\documentclass[%
superscriptaddress,
preprint,
nofootinbib,
 amsmath,amssymb,
 aps,
prd,
]{revtex4}

\usepackage{graphicx}
\usepackage{dcolumn}
\usepackage{bm}
\usepackage{hyperref}
\usepackage{physics}

\usepackage{color}

\begin{document}


\title{S-wave kaon-nucleon potentials with all-to-all propagators in the HAL QCD method}

\newcommand{\Kyoto}{
Center for Gravitational Physics, 
Yukawa Institute for Theoretical Physics, Kyoto University, Kitashirakawa Oiwakecho, Sakyo-ku, 
Kyoto 606-8502, Japan}

\newcommand{\Riken}{RIKEN Nishina Center, Wako 351-0198, Japan}

\author{Kotaro~Murakami}
 \affiliation{\Kyoto}
 \affiliation{\Riken}
\author{Yutaro~Akahoshi}
   \affiliation{\Kyoto}
   \affiliation{\Riken}
\author{Sinya~Aoki}
   \affiliation{\Kyoto}



\collaboration{HAL QCD Collaboration}

\date{\today}


\begin{abstract}
In this paper, employing an all-to-all quark propagator technique, we investigate the kaon-nucleon interactions in lattice QCD. We calculate the S-wave kaon-nucleon potentials at the leading order in the derivative expansion in the time-dependent HAL QCD method, using (2+1)-flavor gauge configurations at the lattice spacing $a \approx 0.09$~fm on $32^3 \times 64$ lattices and the pion mass $m_{\pi} \approx 570$~MeV. We take the one-end trick for all-to-all propagators, which allows us to put the zero momentum hadron operators at both source and sink and to smear quark operators at the source. We find the stronger repulsive interaction in the $I=1$ channel than in the $I=0$. The phase shifts obtained by solving the Schr\"{o}dinger equations with the potentials qualitatively reproduce the energy dependence of the experimental phase shifts, and have the similar behavior to the previous results from lattice QCD without all-to-all propagators. Our study demonstrates that the all-to-all quark propagator technique with the one-end trick is useful to study interactions for meson-baryon systems in the HAL QCD method, so that we will apply it to meson-baryon systems which contain quark-antiquark creation/annihilation processes in our future studies.

\end{abstract}


                   
\preprint{YITP-20-62, RIKEN-QHP-477}
\maketitle


\section{Introduction}

Studying hadronic resonances including exotic hadrons from hadron scatterings is one of the current important issues in lattice QCD. There are two available methods to analyze two-hadron scatterings in lattice QCD. One is the finite volume method~\cite{Luscher:1990ux, Rummukainen:1995vs}, which relates two particle energies measured on finite volume in lattice QCD to scattering phase shifts between two hadrons. The other is the HAL QCD method~\cite{Ishii:2006ec, Aoki:2009ji,HALQCD:2012aa}, where we extract potentials for two hadrons from the NBS wave functions directly and derive the scattering phase shifts by solving the Schor\"{o}dinger equations with the potentials. Since it is very hard to isolate a particular eigenstate of two baryons from other elastic eigenstates, a reliable extraction of the two baryon energy is a challenge in the finite volume method~\cite{Iritani:2016jie}. On the other hand, the HAL QCD method can reduce this difficulty by the time dependent method, which can utilize all elastic scattering states to construct the potential~\cite{HALQCD:2012aa}. Because of this advantage, various kind of the studies on baryon-baryon scatterings in the HAL QCD method have been performed so far~\cite{Ishii:2006ec, Aoki:2009ji, Ishii:2009zr, Inoue:2010hs, Inoue:2010es, Inoue:2011ai, HALQCD:2012aa, Murano:2013xxa, Etminan:2014tya, Doi:2017zov, Gongyo:2017fjb, Miyamoto:2017tjs, Iritani:2018zbt, Iritani:2018sra, Sasaki:2019qnh}.

In nature, hadronic resonances are often observed in the systems which allow quark-antiquark pair creation/annihilation processes, absent in baryon-baryon systems. Studying such systems requires all-to-all quark propagators, which are too time-consuming to calculate exactly, and therefore are evaluated approximately in lattice QCD. Light mesonic resonances such as the $\rho$ resonance have been well investigated from the meson-meson scatterings using all-to-all propagators in the finite volute method~\cite{Briceno:2017max, Alexandrou:2017mpi, Werner:2019hxc}, while such studies have just begun in the HAL QCD method~\cite{Kawai:2017goq,Akahoshi:2019klc}.

In the case of baryonic resonances in meson-baryon systems, there are very few results in the finite volume method~\cite{Lang:2012db, Leskovec:2018lxb,Andersen:2017una, Paul:2018yev}, while none in the HAL QCD method. In the finite volume method, the $\Delta(1232)$ resonance has been studied by the analysis of the $I=3/2$ P-wave pion-nucleon scatterings~\cite{Andersen:2017una, Paul:2018yev}, whose results however have both the statistical and systematic uncertainties probably due to the existence of the baryon. To enhance the reliability of lattice QCD studies on baryonic resonances in meson-baryon systems such as $\Delta(1232)$, it is thus important to investigate such systems in the HAL QCD method, which may avoid a part of the problems in the systems including baryons.

As a first step toward the ambitious goal to reproduce properties of baryonic resonances such as $\Delta(1232)$ from meson-baryon scatterings in lattice QCD, we investigate the $S=+1$ S-wave kaon-nucleon interactions in the HAL QCD method in this study. Although the kaon-nucleon systems allow no quark-antiquark pair creation/annihilation, we employ the all-to-all quark propagator technique with the one-end trick~\cite{Foster:1998vw, McNeile:2002fh}, which will be anyhow necessary for meson-baryon systems with baryonic resonances such as $\Delta(1232)$. All-to-all propagators allow us to use the hadron interpolating operators with zero momenta at both source and sink together with the smeared quark operators at the source. 

For the kaon-nucleon scatterings, several calculations of the phase shifts or the scattering lengths in the finite volume method have been performed~\cite{Fukugita:1994ve, Meng:2003gm, Torok:2009dg, Detmold:2013gua, Detmold:2015qwf}, while there is one previous analysis in the HAL QCD method with the wall source~\cite{Ikeda:2011qm}.

In addition, we focus on the existence of the pentaquark $\Theta^{+}(1540)$, which has been reported by LEPS Collaboration at SPring-8~\cite{Nakano:2003qx}. Although numbers of experimental and theoretical studies have been performed to search for $\Theta^{+}(1540)$, its existence has not been confirmed yet and it is a still controversial issue~\cite{Liu:2014yva}. Since the S-wave kaon-nucleon systems ($I(J^{P})=0(1/2^{-})$ and $1(1/2^{-})$) are the candidates for the channels of $\Theta^{+}(1540)$, we search signals suggesting the existence of $\Theta^{+}(1540)$ in this study.

This paper is organized as follows. In Sec.~\ref{HALQCDmethod}, we briefly review the HAL QCD method. In Sec.~\ref{KN4ptfunc}, we introduce the 4-point correlation functions for the S-wave kaon-nucleon systems and show the quark contraction diagrams. In Sec.~\ref{setup}, the numerical setup is explained. In Sec.~\ref{results}, we present numerical results for the leading-order potentials, and extract the S-wave kaon-nucleon scattering phase shifts, which are compared with experimental data as well as the previous results from lattice QCD. Sec.~\ref{conclusion} is devoted to the conclusion of this paper. In Appendix~\ref{oetsection}, we explain the one-end trick technique.

\section{HAL QCD method}\label{HALQCDmethod}

The NBS wave function at the Euclidean time $t$ is defined as
\begin{equation}\label{nbswf}
\Psi^{W}(\vb{r},t) 
= \Psi^{W}(\vb{r}) \ e^{-Wt} =\bra{0} O_{1}(\vb{x+r} ,t) O_{2}(\vb{x} ,t) \ket{1,2,W},
\end{equation}
where $\ket{0}$ is the vacuum state in QCD, $O_{i}(\vb{x} ,t) \ (i=1,2)$ is the hadron sink operator at $(\vb{x},t)$ and $\ket{1,2,W}$ is the two-hadron states with the energy $W=\sqrt{k^2+m_{1}^2}+\sqrt{k^2+m_{2}^2}$. If we consider the energy region where only the elastic scattering occurs, the NBS wave function satisfies the Schr\"{o}dinger equation for free particles in the limit $|\vb{r}| \to \infty$ as
\begin{equation}
\Big( \frac{k^2}{2\mu}-H_{0}\Big) \Psi^{W}(\vb{r}) \simeq 0,
\end{equation}
where $\mu=m_{1}m_{2}/(m_{1}+m_{2})$ is the reduced mass and $H_{0}=-\nabla^2/2\mu$ is the free part of Hamiltonian. Furthermore, the asymptotic behavior of the $l$-th partial-wave of the NBS wave function reads~\cite{Lin:2001ek, Aoki:2005uf, Aoki:2009ji, Aoki:2013cra}
\begin{eqnarray}
\Psi^{l,W}(r) \underset{|\vb{r}|\to \infty}{\propto} \frac{\sin(kr-\frac{l}{2}\pi + \delta^{l}(k))}{kr} e^{i\delta^{l}(k)},
\end{eqnarray}
where $\delta^{l}(k)$ is the phase shift for the two-hadron scattering.

In the HAL QCD method~\cite{Ishii:2006ec, Aoki:2009ji}, we extract a potential for two hadrons from the NBS wave function as
\begin{equation}
 \int d^3r' \ U(\vb{r},\vb{r}') \Psi^{W}(\vb{r}') 
 = \Big( \frac{k^2}{2\mu} - H_{0} \Big) \Psi^{W}(\vb{r}),
\end{equation}
where the potential $U(\vb{r},\vb{r}')$ is energy-independent and non-local. In addition, $U(\vb{r},\vb{r}')$ can be defined only below the inelastic threshold $W<W_{th}$, and depends on hadron operators we choose.

In lattice QCD, we calculate the 4-point correlation function defined as
\begin{eqnarray}
F(\vb{r},t) = \bra{0}   O_{1}(\vb{x+r} , t+t_{0}) O_{2}(\vb{x} , t+t_{0}) \ \bar{\mathcal{J}} (t_{0}) \ket{0},
\end{eqnarray}
where $\bar{\mathcal{J}} (t_{0})$ is the source operator which creates scattering states $\ket{1,2,W}$. For large $t$, we can extract the NBS wave function for the two hadron state with the lowest energy $W_{0}$ as
\begin{eqnarray}
F(\vb{r},t) 
\underset{t \to \infty}{\simeq}  \bra{1,2, W_{0}} \bar{\mathcal{J}} (0) \ket{0} \Psi^{W_{0}}(\vb{r}) \ e^{-W_{0}t}.
\end{eqnarray}

For the systems including baryons, it is difficult to isolate the ground state from the excited states, since large statistical fluctuations of the 4-point correlation function prevent us from taking large enough $t \gg 1/\Delta E$, where $\Delta E$ is the energy gap between the ground state and the first excited states.

To overcome this difficulty in the kaon-nucleon system, we employ the time-dependent HAL QCD method~\cite{HALQCD:2012aa}, by which we can derive the potential even when there exist contributions from the elastic excited states in the 4-point correlation function. In this method, we use the R-correlator defined as
\begin{eqnarray}
R(\vb{r},t) 
= \frac{F(\vb{r},t)}{C_{1}(t)C_{2}(t)},
\end{eqnarray}
where $C_{1}(t)$ and $C_{2}(t)$ are the 2-point correlation functions for hadron 1 and 2, respectively. When $t$ is large enough to suppress inelastic contributions, the R-correlator reads
\begin{eqnarray}\label{propofRcorr}
\begin{aligned}
R(\vb{r},t)
\simeq \sum_{n} \frac{  A_{n} \Psi^{W_{n}}(\vb{r}) \ e^{-W_{n}t} }{e^{-m_{1}t} e^{-m_{2}t}}  
 = \sum_{n}   A_{n} \Psi^{W_{n}}(\vb{r}) \ e^{-\Delta  W_{n}t},
\end{aligned}
\end{eqnarray}
where $A_{n}$ is the factor independent of $\vb{r}$ and $\Delta  W_{n} = W_{n}-m_{1}-m_{2}$ is the energy from the threshold.

For $m_{1} \neq m_{2}$, $\Delta W_{n}$ and $k_{n}^{2}$ are related as
\begin{eqnarray}
\frac{k_{n}^2}{2\mu} =  \frac{1+3\delta^2}{8\mu}\Delta W_{n}^2+\Delta W_{n}+\order{\Delta W_{n}^3}
\end{eqnarray}
for small $\Delta W_{n}$ and $k^2_{n}$, where $\delta = (m_{1}-m_{2})/(m_{1}+m_{2})$. Thus, the R-correlator satisfies
\begin{eqnarray}
\Big(  \frac{1+3\delta^2}{8\mu}\pdv[2]{t} -\pdv{t} -H_{0} \Big)  R(\vb{r},t)
=  \int d^3r' \ U(\vb{r},\vb{r}')   R(\vb{r}',t) +\order{\Delta W_{n}^3}.
\end{eqnarray}
In this paper, we approximate $U(\vb{r},\vb{r}')$ by the leading order of the derivative expansion as
\begin{eqnarray}
U(\vb{r},\vb{r}') \approx V^{\textrm{LO}}_{0}(r)\delta^{(3)}(\vb{r}-\vb{r}'),
\end{eqnarray}
in which $V^{\textrm{LO}}_{0}(r)$ can be extracted as
\begin{eqnarray}
V^{\textrm{LO}}_{0}(r) = \frac{1}{R(\vb{r},t)} \Big(  \frac{1+3\delta^2}{8\mu}\pdv[2]{t} -\pdv{t} -H_{0} \Big)  R(\vb{r},t). \label{timedepnoneqmass}
\end{eqnarray}

\section{Interpolating operators and 4-point correlation functions for S-wave kaon-nucleon systems}\label{KN4ptfunc}

The interpolating operators are defined for kaons as
\begin{eqnarray}
\begin{aligned}
K^{+}(x) &=& i\bar{s}(x)\gamma_{5}u(x), \quad K^{-}(x) = i\bar{u}(x)\gamma_{5}s(x), \\
K^{0}(x) &=&  i\bar{s}(x)\gamma_{5}d(x), \quad \bar{K}^{0}(x) =  i\bar{d}(x)\gamma_{5}s(x),
\end{aligned}
\end{eqnarray}
and for nucleons as
\begin{eqnarray}
\begin{aligned}
N(x)_{\alpha} &=& \epsilon_{abc}q_{a,\alpha}(x)(u^{{\textrm T}}_{b}(x) C\gamma_{5}d_{c}(x)), \\
\bar{N}(x)_{\alpha} &=& -\epsilon_{abc}\bar{q}_{a,\alpha}(x)(\bar{u}_{b}(x) C\gamma_{5}\bar{d}^{{\textrm T}}_{c}(x)),
\end{aligned}
\end{eqnarray}
where $N=(p,n)$ and $q=(u,d)$.

The 4-point correlation function for the S-wave kaon-nucleon system in the $I=1$ channel is defined as
\begin{eqnarray}\label{I14ptfunc}
 F^{I=1}_{\alpha \beta}({\bf r},t;\vb{z}_{0}, t_{0}) =  \langle J^{I=1}_{\alpha}({\bf r},t+t_0) \bar{J}^{I=1}_{\beta}(\vb{z}_{0},t_0) \rangle,
\end{eqnarray}
where $J^{I=1}$ is the sink operator given by
\begin{eqnarray}\label{I1sinkop}
\begin{aligned}
J^{I=1}_{\alpha}({\bf r},t+t_0) 
&= \sum_{{\bf x}} K^{+}(\underbrace{{\bf r+x},t+t_0}_{x_1}) \ p_{\alpha}(\underbrace{{\bf x},t+t_0}_{x_2}) \\
&=i\sum_{{\bf x}}(\bar{s}(x_1)\gamma_{5}u(x_1)) \ (\epsilon_{abc}u_{a,\alpha}(x_2)(u^{{\textrm T}}_{b}(x_2) C\gamma_{5}d_{c}(x_2)) 
\end{aligned}
\end{eqnarray}
and $\bar{J}^{I=1}$ is the source operator by
\begin{eqnarray}\label{I1srcop}
\begin{aligned}
\bar{J}^{I=1}_{\beta}(\vb{z}_{0},t_0) &= \sum_{{\bf y}} K^{-}(\underbrace{{\bf y},t_0}_{y})  \bar{p}_{\beta}(\underbrace{{\bf z}_{0},t_0}_{z_{0}}) \\
&=-i\sum_{{\bf y}} (\bar{u}(y)\gamma_{5}s(y)) \ (\epsilon_{a'b'c'}\bar{u}_{a',\beta}(z_{0})(\bar{u}_{b'}(z_{0}) C\gamma_{5}\bar{d}^{{\textrm T}}_{c'}(z_{0})).
\end{aligned}
\end{eqnarray}
In order for $F^{I=1}$ to have the strong overlap with the S-wave ground state, we choose both $K^{-}$ and $\bar{p}$ operators with zero momenta. However, once the coordinate ${\bf x}$ in Eq.(\ref{I1sinkop}) is summed over to fix the total momentum of the system to be zero, only one of the two operators has to be projected thanks to the momentum conservation of the system. We therefore sum over ${\bf y}$ to project $K^{-}$ onto zero momentum and keep the coordinate ${\bf z}_{0}$ without summation. 

\begin{figure}
	\centering
	\includegraphics[width=12cm]{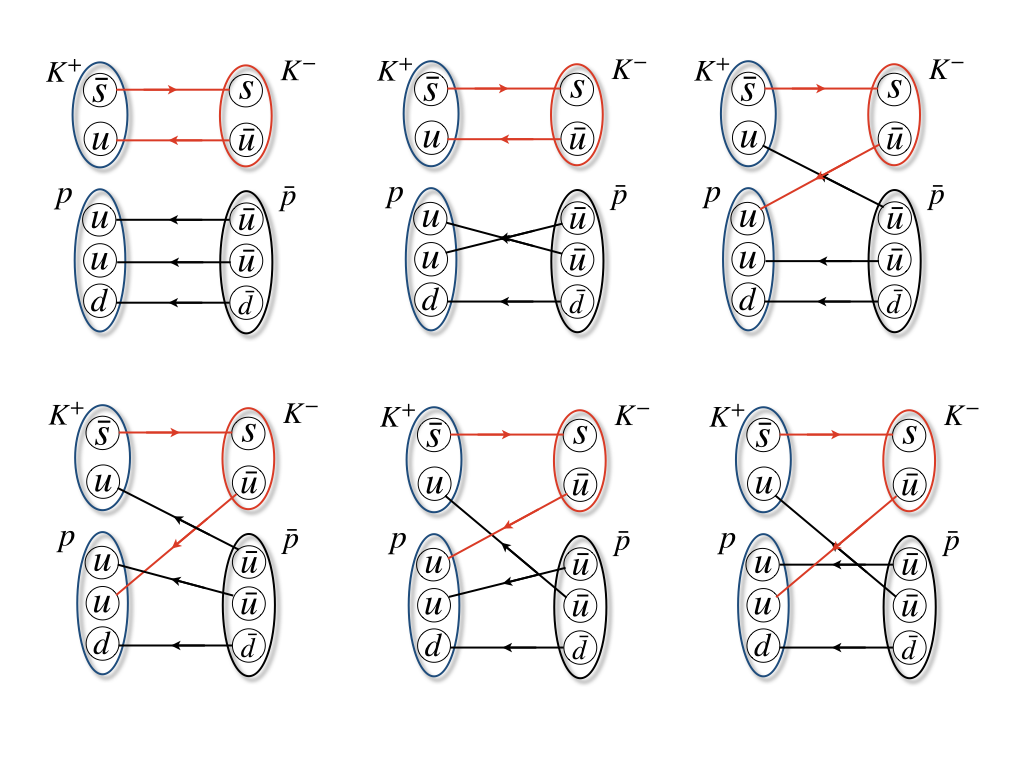}
	\caption{Quark contraction diagrams for the kaon-nucleon system in the $I=1$ channel. Operators on the left-hand side and the right-hand side are sink and source operators, respectively. Red lines represent all-to-all propagators and black lines show point-to-all propagators.}
	\label{NKI1cont}
\end{figure}

Fig.~\ref{NKI1cont} shows the quark contraction diagrams for the kaon-nucleon system in the $I=1$ channel, where there is no quark-antiquark pair annihilation/creation. Nevertheless, since we sum over the spatial coordinate of the $K^{-}$ operator (red circles in Fig.~\ref{NKI1cont}), all-to-all propagators are needed for red lines in Fig.~\ref{NKI1cont}. We thus use the one-end trick~\cite{Foster:1998vw, McNeile:2002fh}, which is the technique to evaluate approximately the products of two all-to-all propagators such as red lines connected by red circles in Fig.~\ref{NKI1cont}. We briefly explain the one-end trick in Appendix~\ref{oetsection}.

\begin{figure}
	\centering
	\includegraphics[width=10cm]{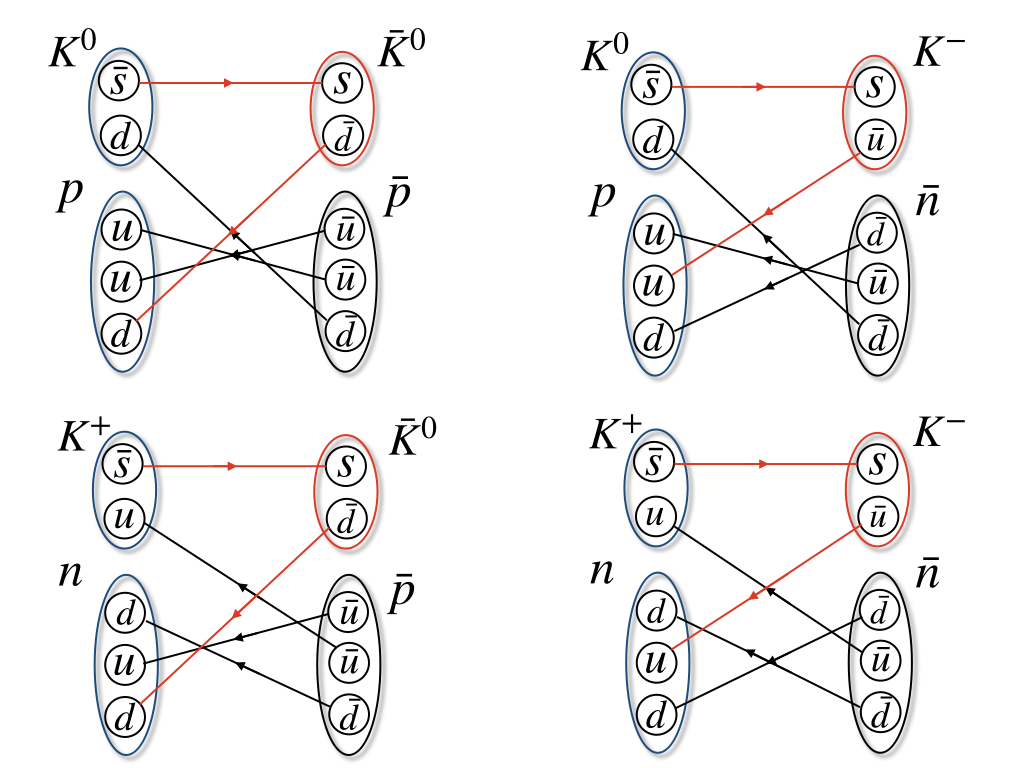}
	\caption{Quark contraction diagrams that appear only for the kaon-nucleon system in the $I=0$ channel. Red lines represent all-to-all propagators and black lines show point-to-all propagators.}
	\label{NKI0cont}
\end{figure}

For the S-wave kaon-nucleon system in the $I=0$ channel, the 4-point correlation function is given by
\begin{eqnarray}
 F^{I=0}_{\alpha \beta}({\bf r},t;\vb{z}_{0},t_{0}) =  \langle J^{I=0}_{\alpha}({\bf r},t+t_0) \bar{J}^{I=0}_{\beta}(\vb{z}_{0},t_0) \rangle \label{I04ptfunc},
\end{eqnarray}
where 
\begin{eqnarray}
\begin{aligned}
J^{I=0}_{\alpha}({\bf r},t+t_0) 
= \sum_{{\bf x}}&(K^{0}(x_{1}) \ p_{\alpha}(x_2)-K^{+}(x_{1}) \ n_{\alpha}(x_2)) \\
=i\sum_{{\bf x}}&[(\bar{s}(x_1)\gamma_{5}d(x_1)) \ (\epsilon_{abc}u_{a, \alpha}(x_2)(u^{{\textrm T}}_{b}(x_2) C\gamma_{5}d_{c}(x_2)) \\
-&(\bar{s}(x_1)\gamma_{5}u(x_1)) \ (\epsilon_{abc}d_{a, \alpha}(x_2)(u^{{\textrm T}}_{b}(x_2) C\gamma_{5}d_{c}(x_2)) ]
\end{aligned}
\end{eqnarray}
and
\begin{eqnarray}
\begin{aligned}
\bar{J}^{I=0}_{\beta}(\vb{z}_{0},t_0) 
= \sum_{{\bf y}} &(\bar{K}^{0}(y) \ \bar{p}_{\beta}(z_{0})-K^{-}(y) \ \bar{n}_{\beta}(z_{0})) \\
=-i\sum_{{\bf y}}&[(\bar{d}(y)\gamma_{5}s(y)) \ (\epsilon_{a'b'c'}\bar{u}_{a', \beta}(z_{0})(\bar{u}_{b'}(z_{0}) C\gamma_{5}\bar{d}^{{\textrm T}}_{c'}(z_{0})) \\
&-(\bar{u}(y)\gamma_{5}s(y)) \ (\epsilon_{a'b'c'}\bar{d}_{a', \beta}(z_{0})(\bar{u}_{b'}(z_{0}) C\gamma_{5}\bar{d}^{{\textrm T}}_{c'}(z_{0})) ]
\end{aligned}
\end{eqnarray}
for $x_{1}=({\bf r+x},t+t_0)$, $x_{2}=({\bf x},t+t_0)$, $y=({\bf y},t_0)$ and $z_{0}=({\bf z}_{0},t_0)$. Because of the isospin symmetry in our calculation, quark contraction diagrams for $F^{I=0}$ consist of those in $F^{I=1}$ with different coefficients plus diagrams shown in Fig.~\ref{NKI0cont}. As is the case for $F^{I=1}$, red circles in Fig.~\ref{NKI0cont} show the kaon operators whose spatial coordinates are summed over and the products of two all-to-all propagators represented as red lines and circles in Fig~\ref{NKI0cont} are calculated using the one-end trick.

\section{Numerical setup}\label{setup}

In our calculation of the kaon-nucleon 4-point correlation functions, we use (2+1)-flavor gauge configurations generated by PACS-CS Collaboration with the improved Iwasaki gauge action and the $\order{a}$-improved Wilson quark action at $\beta = 1.90$ on $32^3 \times 64$ lattice volume~\cite{Aoki:2008sm}, which corresponds to $a\approx0.09$~fm for the lattice spacing. The hopping parameters of the ensemble in our calculation are $\kappa_{u(d)}=0.13727$ and $\kappa_{s}=0.13640$, and the corresponding pion mass is $m_{\pi} \approx 570$~MeV according to Ref.~\cite{Aoki:2008sm}. Kaon and nucleon masses from corresponding 2-point correlation functions calculated in our study give $m_{K}=713(1)$~MeV and $m_{N} = 1405(7)$~MeV, respectively. The periodic boundary condition is imposed in all spacetime directions. We used 400 configurations with 4 sources at different time slices on each configuration. Statistical errors are estimated by the jackknife method with a binsize of 40 configurations.

To the noise vector $\eta(\vb{x})$ for the one-end trick we apply dilution~\cite{Foley:2005ac} for color and spinor components and s2 dilution defined by the following equation
\begin{eqnarray}
\eta^{(s_{dil})}(\vb{x}) =
 \begin{cases}
\eta(\vb{x})	& (x+y+z \equiv s_{dil} \pmod{2}) \\
0			& (x+y+z \equiv s_{dil}+1 \pmod{2})
 \end{cases} 
 , \ s=0,1,
\end{eqnarray}
so that the noise due to $\eta(\vb{x})$ is reduced. We employ the smeared quark source using the smearing function~\cite{Iritani:2016jie} given by
\begin{eqnarray}
 f_{A,B}(\vb{r}) =
 \begin{cases}
A e^{-B|\vb{r}|} & (|\vb{r}|<\frac{L-1}{2}) \\
1 & (|\vb{r}|=0) \\
0 & (|\vb{r}| \geq\frac{L-1}{2})
 \end{cases} \label{smearingfunction}
\end{eqnarray}
in lattice unit, where we take $(A,B)=(1.2, 0.19)$ for up and down quarks and $(A,B)=(1.2, 0.25)$ for strange quarks. Furthermore, to obtain the S-wave NBS wave functions, we project each spinor component of the 4-point correlation functions $F^{I=1(0)}_{\alpha \beta}({\bf r},t;{\bf z}_{0},t_{0})$ onto the A$^{+}_{1}$ representation of the cubic group $O_{h}$, which is the discrete rotational symmetry on the lattice. 

In order to reduce the statistical fluctuations, we apply the all-mode averaging~\cite{Shintani:2014vja} without low mode averages. Using the translational invariance, we calculate the 4-point correlation functions for the nucleon source operator at ${\bf z_{0}}$ with the stopping condition of the Bi-CGSTAB solver~$||D\psi - s||/||s||< \epsilon$ set to $\epsilon=10^{-12}$ as the exact ones, followed by the approximated evaluations with $\epsilon=10^{-4}$ at 8 different source points, given by ${\bf z_{0}}+\Delta{\bf z_{0}}$ with $\Delta{\bf z_{0}} = (0,0,0), (0,0,L/2), \cdots, (L/2,L/2,L/2)$, where $L=32$ is the lattice spatial extension. The location ${\bf z_{0}}$ is randomly chosen on each configuration to reduce a possible bias due to the inexact Bi-CGSTAB solver.

\section{S-wave kaon-nucleon potentials and phase shifts}\label{results}
 
\subsection{Leading-order potentials}
 
\begin{figure}
    \begin{center}
        \includegraphics[width=0.49\textwidth]{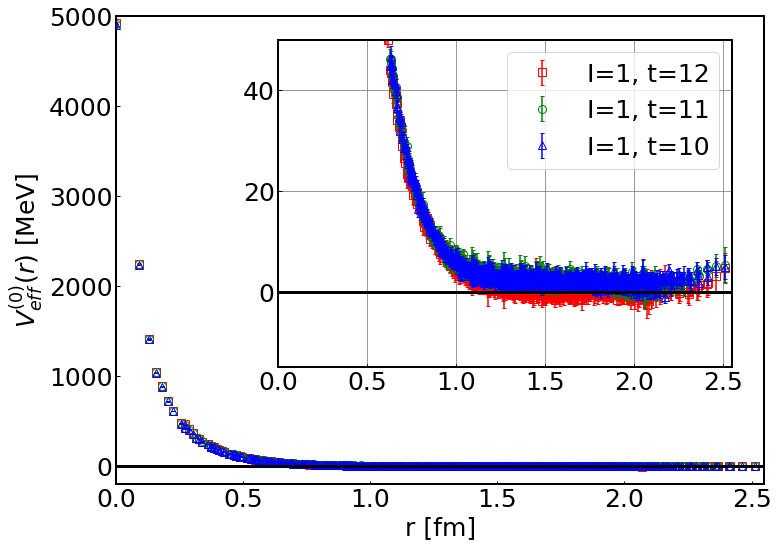}
	    \includegraphics[width=0.49\textwidth]{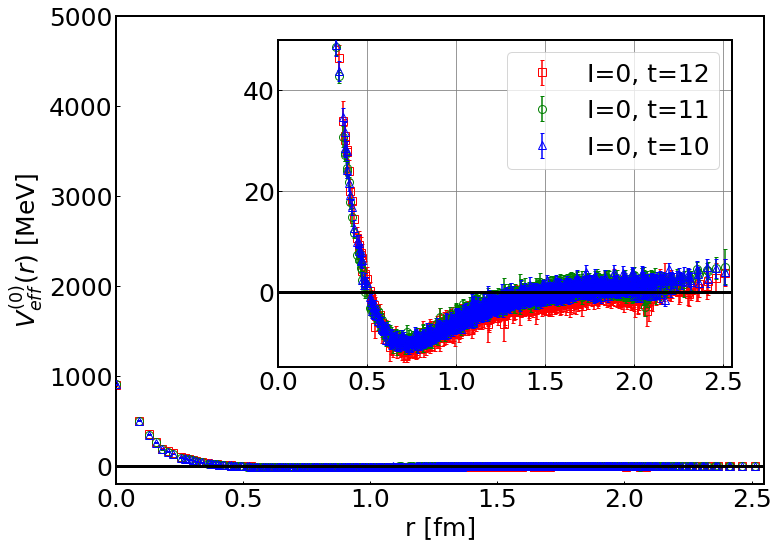}
	    \caption{The leading-order potentials $V_{0}(r)$ for kaon-nucleon system in $I=1$ (Left) and $I=0$ (Right) channel at $t=10-12$.}
	    \label{potential}    
    \end{center}
\end{figure}
\begin{figure}
    \begin{center}
        \includegraphics[width=0.49\textwidth]{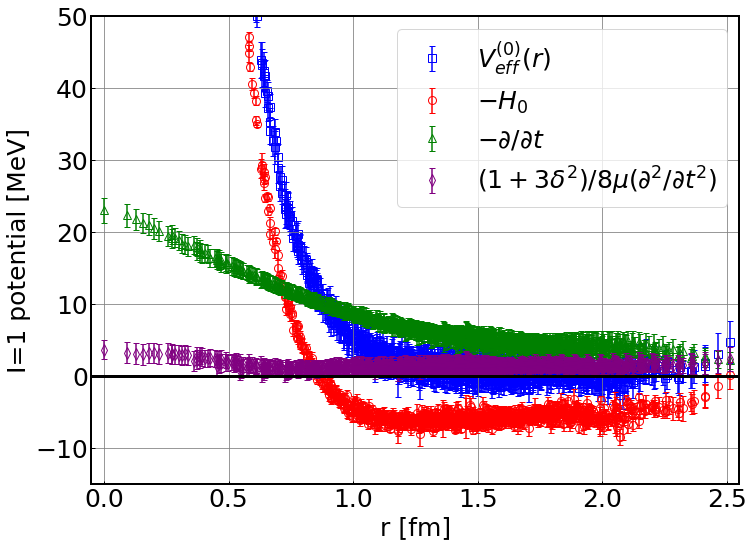}
	    \includegraphics[width=0.49\textwidth]{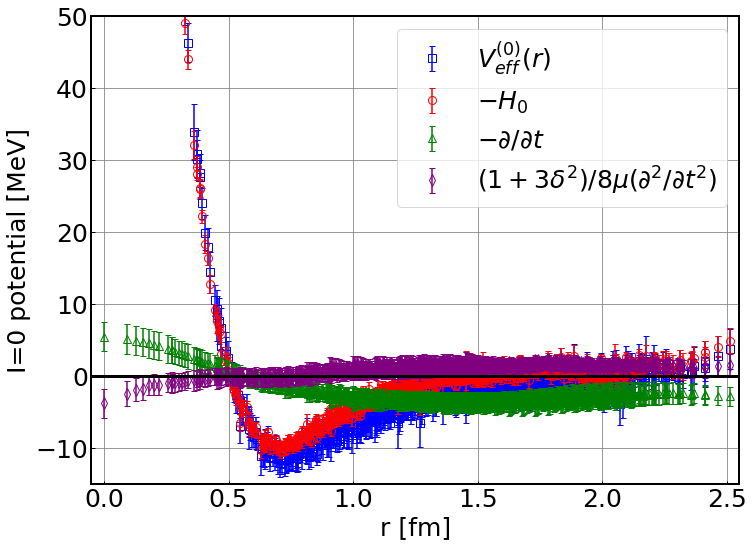}
	    \caption{The potentials and the terms in~Eq.(\ref{timedepnoneqmass}) at $t=12$ for $I=1$ (Left) and $I=0$ (Right).}
	    \label{potdec}    
    \end{center}
\end{figure}
 
In Fig.~\ref{potential}, we present the leading-order potentials for the S-wave kaon-nucleon systems at $t=10-12$, which is the imaginary-time region where the effective masses of both the kaon and nucleon 2-point correlation functions are saturated and the signals of the potentials do not suffer from the noises so much. Both the $I=1$ and $I=0$ potentials show very weak $t$-dependences, which indicates that not only contaminations from inelastic states but also contributions from higher-order terms in the derivative expansion are well under control at $t=10$. In addition, both potentials become zero within errors at long distances ($1.5<r<2.5$ fm), indicating the absence of significant inelastic contributions. While the $I=1$ potential in Fig.~\ref{potential} (Left) is repulsive at all distances with the repulsive core at short distances ($0<r<0.5$ fm), the $I=0$ potential in Fig.~\ref{potential} (Right) has both the repulsive core at short distances and the shallow attractive pocket with the depth of about $10$~MeV at the middle distances ($0.5<r<1.5$ fm). The repulsive core of the $I=1$ potential at the short distance is much stronger than that of the $I=0$ potential.

Fig.~\ref{potdec} represents the potentials and their breakups into three terms in~Eq.(\ref{timedepnoneqmass}) at $t=12$. The 2nd derivative terms (purple diamonds) are small at the middle and long distances. Indeed we find that the potentials with and without these terms give almost identical results on the phase shifts. Therefore, the $\order{\Delta W^3}$ relativistic corrections in Eq.(\ref{timedepnoneqmass}), absent in our calculation, are expected to be further reduced.

\subsection{Phase shifts}

\begin{figure}
    \begin{center}
        \includegraphics[width=0.49\textwidth]{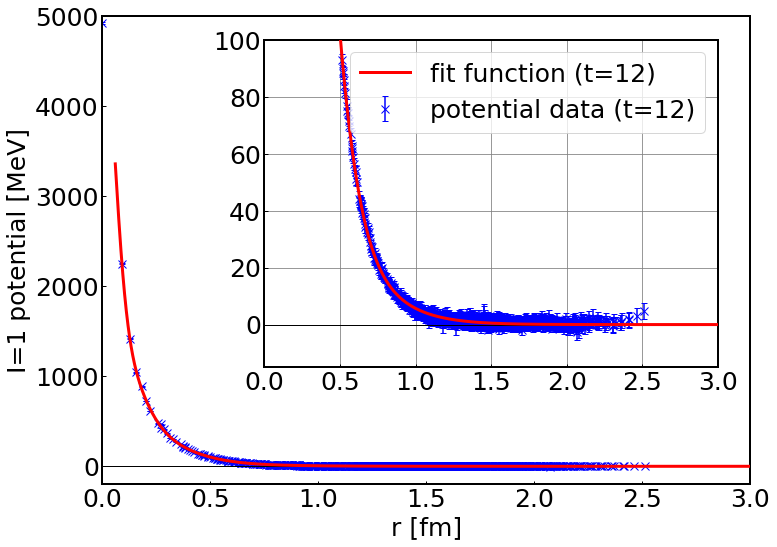}
	    \includegraphics[width=0.48\textwidth]{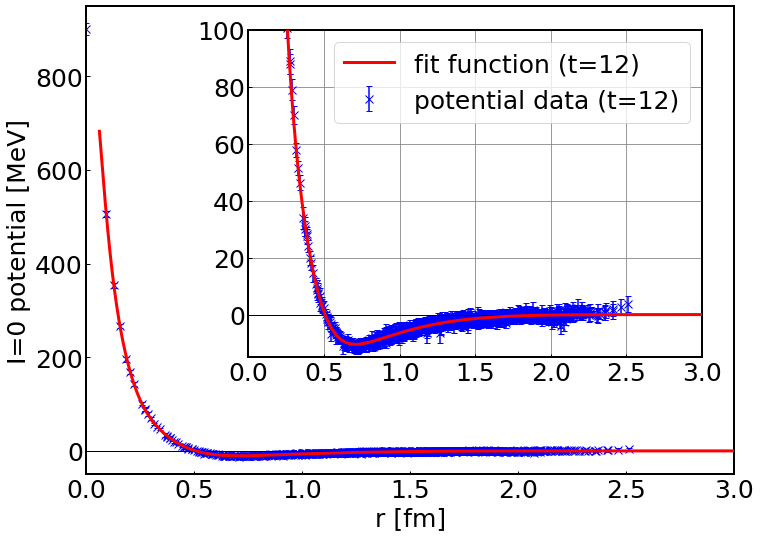}
	    \caption{The fitted potentials (Red lines) for $I=1$ (Left) and $I=0$ (Right) at $t=12$. Blue crosses show the potential data at $t=12$.}
	    \label{potentialfit}    
    \end{center}
\end{figure}

We fit the potentials at $t=12$ by the sum of four Gaussians given by
\begin{eqnarray}
V(r) = a_{0}e^{-(r/a_{1})^2}+a_{2}e^{-(r/a_{3})^2}+a_{4}e^{-(r/a_{5})^2}+a_{6}e^{-(r/a_{7})^2}, \label{fitfunc}
\end{eqnarray}
where we assume that $a_{1} < a_{3} < a_{5} < a_{7}$. The fit parameters are listed in Table~\ref{fitparam} and Fig.~\ref{potentialfit} compares the fitted potential and the data at $t=12$.

\begin{table}[b]
\caption{\label{fitparam}%
Fit parameters $a_{i}$ for the $I=1$ and $I=0$ potential data at $t=12$. $\chi^2/dof=0.65(0.19)$ for $I=1$ and $\chi^2/dof=0.27(0.09)$ for $I=0$.
}
\begin{ruledtabular}
\begin{tabular}{l|ccccccccc}
$I$    &
$a_{0}$ [MeV]&
$a_{1}$ [fm]&
$a_{2}$ [MeV]&
$a_{3}$ [fm]&
$a_{4}$ [MeV]&
$a_{5}$ [fm]&
$a_{6}$ [MeV]&
$a_{7}$ [fm]&
 \\
\colrule
$1$   &  3180(193)& 0.078(0.002)& 
           1185(17)& 0.176(0.024)&
            501(97)& 0.357(0.075)&
            61(111)& 0.699(0.357) \\
$0$   & 352(121)&  0.079(0.012)&
        381(86)&  0.151(0.029)&
        189(34)&  0.325(0.029)&
        -23(8)&  0.898(0.218)& \\
\end{tabular}
\end{ruledtabular}
\end{table}

\begin{figure}
	\begin{center}
	\includegraphics[width=0.86\textwidth]{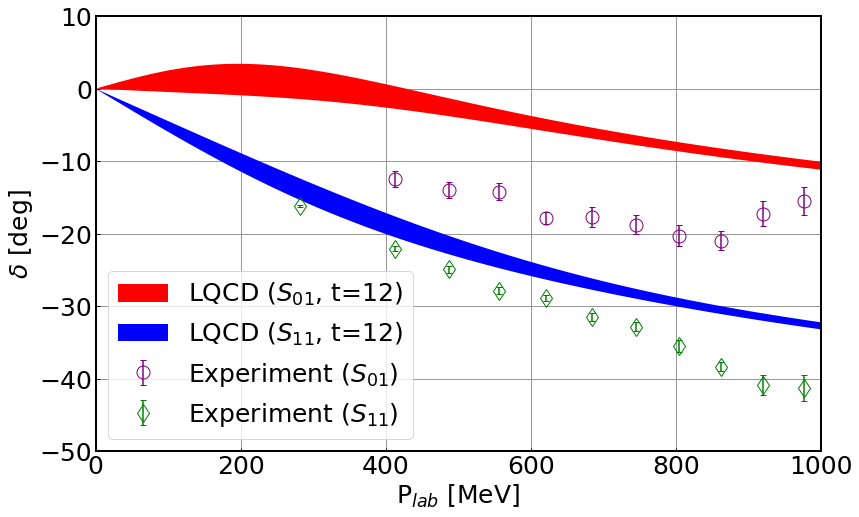}
	\caption{The S-wave kaon-nucleon scattering phase shifts for $I=1$ (blue band) and $I=0$ (red band) in this study. The experimental results~\cite{SAIDKN} are plotted by green diamonds for $I=1$ and by purple circles for $I=0$ in this figure.}
	\label{phase}
	\end{center}
\end{figure}

We solve the Schr\"{o}dinger equations in the radial direction as
\begin{eqnarray}
 -\frac{1}{2\mu}\Big[\frac{1}{r}\dv{r}(r\dv{r}) + \frac{1}{r}\dv{r} - \frac{l(l+1)}{r^2}\Big]\psi^{l,E}_{R}(r) + V(r)\psi^{l,E}_{R}(r) = E\psi^{l,E}_{R}(r),
\end{eqnarray}
where $V(r)$ is the fitted potential in~Eq.(\ref{fitfunc}) and the angular momentum is set to $l=0$ for the S-wave channel. From the solutions, we extract the phase shifts for the S-wave kaon-nucleon scattering.

Fig.~\ref{phase} shows the phase shifts as functions of the kaon momentum in the laboratory frame $P_{lab}$. We find that the $I=1$ phase shift is negative in all the energy region due to the repulsive potential at all distances in Fig.~\ref{potential} (Left). On the other hand, the $I=0$ phase shift becomes slightly positive in the low energy region ($0 < P_{lab} < 400$~MeV), reflecting the small attractive pocket seen in Fig~\ref{potential} (Right), and then turns into negative in the higher energy region ($400 < P_{lab} < 1000$~MeV) due to the repulsion at shorter distance. The overall behavior of the scattering phase shifts obtained through the potentials in this study is roughly consistent with that of the experimental results~\cite{SAIDKN}, plotted by green diamonds ($I=1$) and purple circles ($I=0$) in this figure. In both lattice QCD and experiment, the scattering phase shifts are almost repulsive and decrease as $P_{lab}$ increases, and the $I=1$ channel is more repulsive than the $I=0$ channel. At the quantitative level, however, the phase shifts from lattice QCD are less repulsive than the experimental results, probably due to the difference of the pion mass, $570$~MeV in this study and $140$~MeV in Nature.

For the comparison to the previous theoretical studies, in the constituent quark model of hadrons~\cite{PhysRevC.49.1166}, the similar repulsive behavior in the $I=1$ and $I=0$ channels can be seen, while the small attraction in the $I=0$ channel has not been predicted. The phase shift results in the (2+1)-flavor lattice QCD calculations in the finite volume method at the lighter pion masses~\cite{Torok:2009dg, Detmold:2013gua, Detmold:2015qwf} have similar but more negative values than our results. These suggest that the attractive pocket in the $I=0$ potential may disappear when we perform the calculation at the lighter quark mass, which is an interesting point to see in future studies. Furthermore, the $I=1$ phase shift in this study and in the previous study by the time-dependent HAL QCD method with the wall source at $m_{\pi} \approx 700$~MeV~\cite{Ikeda:2011qm}, which is slightly heavier than that in this study\footnote{The study with the wall source at $m_{\pi} \approx 570$~MeV~\cite{Ikeda:up} has not been published, but we have confirmed that the phase shift results are almost the same as those at $m_{\pi} \approx 700$~MeV.}, are found to agree well in the low energy region but slightly differ by about $3$ degrees in the high energy region. This difference may be explained by the difference of the quark masses or by the systematic uncertainty due to the higher-order contributions in the derivative expansion. Even if this small difference really represents an effect of quark masses, the $I=1$ kaon-nucleon scattering phase shift seems rather insensitive to quark masses in this range of the pion masses.

Finally, we should mention that there exists no resonances or bound states in both channels. This implies that $\Theta^{+}(1540)$ does not exist in the S-wave kaon-nucleon systems ($I(J^{P})=0(1/2^{-})$ and $1(1/2^{-})$ channels) for the quark masses corresponding to $m_{\pi} \approx 570$~MeV.

\section{Conclusion}\label{conclusion}

In this paper, we have calculated the leading-order potentials in the derivative expansion using the time-dependent HAL QCD method for the S-wave kaon-nucleon systems at $m_{\pi}\approx 570$~MeV, and extracted the scattering phase shifts by solving the Schr\"{o}dinger equations with the fitted potentials. We have employed all-to-all propagators with the one-end trick to use the kaon sources with zero momenta as well as the smeared quark sources. In addition, the all-mode averaging has been performed in our calculation. 

We have found that both the $I=1$ and $I=0$ potentials have repulsive cores while the $I=1$ potential is more repulsive than the $I=0$ potential. The scattering phase shifts in this study qualitatively reproduce the energy dependences of the experimental data, and are consistent with the previous results from lattice QCD as well. These results suggest that all-to-all propagator technique with the one-end trick works well to investigate meson-baryon interactions in the HAL QCD method. This work is the first step toward the studies on baryonic resonances in the HAL QCD method, in particular, the investigation of the $\Delta(1232)$ resonance from the $I(J^{P})=3/2(3/2^{+})$ pion-nucleon scatterings.

Furthermore, we have found that there appear no resonances or bound states corresponding to $\Theta^{+}(1540)$ in the behaviors of the kaon-nucleon scattering phase shifts in the $J^{P}=1/2^{-}$ with $I=0,1$ channels at the quark mass in this setup.

\section{Acknowledgements}
We use lattice QCD code of Bridge++ ~\cite{bridge, Ueda:2014rya} and our numerical calculation has been performed on Cray XC40 at Yukawa Institute for Theoretical Physics (YITP) in Kyoto University and HOKUSAI BigWaterfall at RIKEN. We thank the PACS-CS Collaboration for providing us their gauge configurations. This work is supported in part by the Grant-in-Aid of the Japanese Ministry of Education, Sciences and Technology, Sports and Culture (MEXT) for Scientific Research (Nos. JP16H03978, JP18H05236), by a priority issue (Elucidation of the fundamental laws and evolution of the universe) to be tackled by using Post ``K" Computer, and by Joint Institute for Computational Fundamental Science (JICFuS). This work is also supported in part by Program for Promoting Researches on the Supercomputer Fugaku” (Simulation for basic science: from fundamental laws of particles to creation of nuclei). Y.A. is supported in part by the Japan Society for the Promotion of Science (JSPS). We thank other members of the HAL Collaboration for fruitful discussions.
 
\appendix

\section{One-end trick}\label{oetsection}
\begin{figure}
   \centering
   \includegraphics[width=0.8\textwidth]{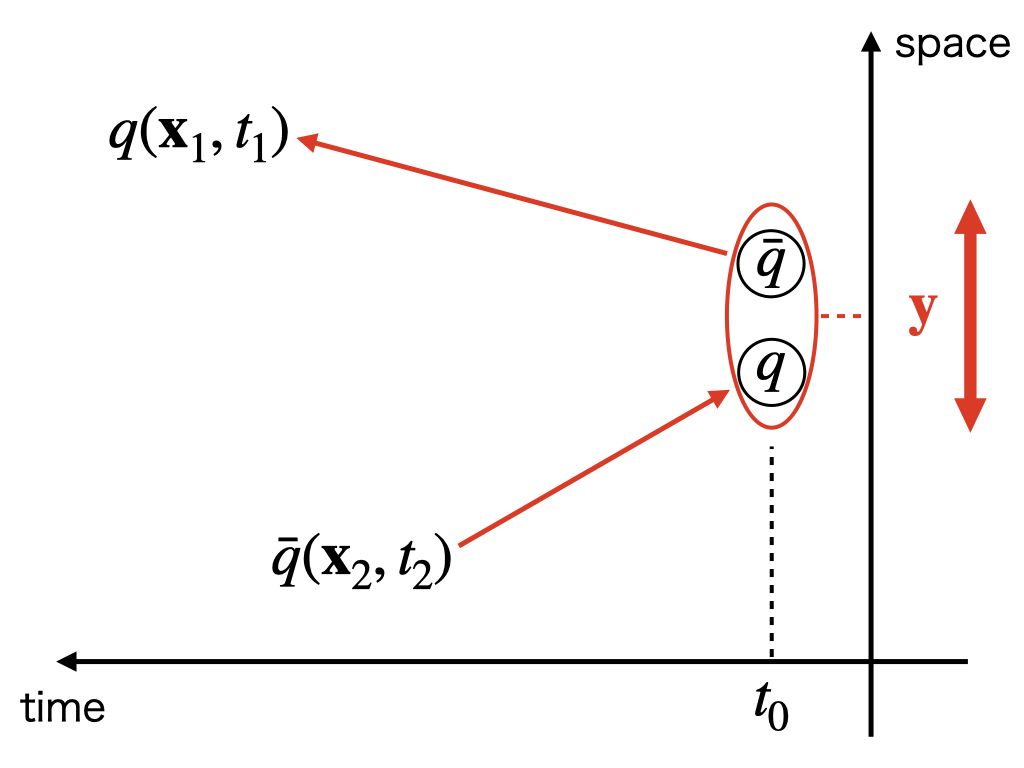}
   \caption{Typical quark contraction diagram to which the one-end trick can be applied.}
   \label{oet}
\end{figure}
In this appendix, we explain the one-end trick~\cite{Foster:1998vw, McNeile:2002fh}, which is one of the technique for the all-to-all propagator calculations. At first, we introduce the noise vector at a time slice defined as
\begin{equation}\label{defofnoisevec1}
\eta^{[r](t_{0})}_{a, \alpha}({\bf x},t_{x}) = \delta_{t_{x},t_{0}}\Xi_{a, \alpha}^{[r]}({\bf x}),
\end{equation}
where $\Xi_{a, \alpha}^{[r]}({\bf x})$ is the time-independent noise vector which satisfies the following equation
\begin{equation}\label{defofnoisevec2}
\langle \Xi_{a, \alpha}^{[r]}({\bf x}) \Xi^{[r]\dagger}_{b, \beta}({\bf y}) \rangle
\equiv \lim_{N \to \infty} \frac{1}{N} \sum_{r=1}^{N}\Xi_{a, \alpha}^{[r]}({\bf x}) \Xi^{[r]\dagger}_{b, \beta}({\bf y}) 
= \delta_{a,b} \delta_{\alpha, \beta} \delta_{{\bf x},{\bf y}}.
\end{equation}
Z2 or Z4 noise is often used to generate $\Xi^{[r]}$.

We consider the product of two propagators given by
\begin{equation}\label{oetformula}
\sum_{{\bf y}} G({\bf x}_1, t_{1};{\bf y},t_0)\Gamma G({\bf y},t_0;{\bf x}_2, t_{2}),
\end{equation}
where $\Gamma$ is some gamma matrix or product of gamma matrices. Here we abbreviate the spinor and color indices. Fig.\ref{oet} represents the quark contraction diagram for Eq.(\ref{oetformula}). In order to calculate Eq.(\ref{oetformula}) for arbitrary $({\bf x}_{1}, t_{1})$ and $({\bf x}_{2}, t_{2})$, the two red lines requires all-to-all propagators. In the one-end trick, we insert the Kronecker deltas between $G({\bf x}_1, t_{1};{\bf y},t_0)$ and $\Gamma$ in Eq.(\ref{oetformula}) and estimate them stochastically by using Eq.(\ref{defofnoisevec1}) and Eq.(\ref{defofnoisevec2}) as
\begin{eqnarray}
\begin{aligned}\label{oetmethod1}
 &\sum_{{\bf y}}\sum_{{\bf z},t_{z},t_{y}} G({\bf x}_1, t_{1};{\bf z},t_{z})
 (\delta_{t_{z},t_{0}}\delta_{t_{y},t_{0}}\delta_{{\bf z},{\bf y}})
 \Gamma G({\bf y},t_y;{\bf x}_2, t_{2}) \\
 \simeq &\sum_{{\bf y}}\sum_{{\bf z},t_{z},t_{y}} G({\bf x}_1, t_{1};{\bf z},t_{z})
 \Big( \frac{1}{N}\sum_{r=1}^{N}\eta^{[r](t_{0})} ({\bf z},t_{z}) \otimes \eta^{[r](t_{0})\dagger}({\bf y},t_{y}) \Big)\Gamma G({\bf y},t_y;{\bf x}_2, t_{2}). \\
 \end{aligned}
\end{eqnarray}
Using $\gamma_{5}$ hermiticity of $G({\bf y},t_y;{\bf x}_2, t_{2})$, Eq.(\ref{oetmethod1}) reads
\begin{eqnarray}
\begin{aligned}
 &\frac{1}{N}\sum_{r=1}^{N} \Big( \sum_{{\bf z},t_{z}}G({\bf x}_1, t_{1};{\bf z},t_{z})\eta^{[r](t_{0})}({\bf z},t_{z}) \Big) 
 \otimes \Big(\sum_{{\bf y},t_{y}} \eta^{[r](t_{0})\dagger}({\bf y},t_{y})  \Gamma\gamma_{5}G^{\dagger}({\bf y},t_y;{\bf x}_2, t_{2})\gamma_{5}\Big)  \\
 =&\frac{1}{N}\sum_{r=1}^{N} (G \eta^{[r](t_{0})})({\bf x}_1, t_{1}) \otimes\big((G \gamma_{5} \Gamma^{\dagger} \eta^{[r](t_{0})})^{\dagger}({\bf x}_2, t_{2})\gamma_{5}\big).
\end{aligned}
\end{eqnarray}
Therefore, solving $2N$ linear equations
\begin{eqnarray}\label{lineareq}
D\psi^{[r](t_{0})} &=& \eta^{[r](t_{0})}, \\
D\xi^{[r](t_{0})} &=& \gamma_{5} \Gamma^{\dagger} \eta^{[r](t_{0})},
\end{eqnarray}
where $D$ is the Dirac operator, we have
\begin{equation}
\sum_{{\bf y}} G({\bf x}_1, t_{1};{\bf y},t_0)\Gamma G({\bf y},t_0;{\bf x}_2, t_{2}) \simeq \frac{1}{N}\sum_{r=1}^{N} \psi^{[r](t_{0})}({\bf x}_1, t_{1}) \otimes (\xi^{[r](t_{0})\dagger}({\bf x}_2, t_{2})\gamma_{5}).
\end{equation}
When $\Gamma = \gamma_{5}$, we only have to solve $N$ linear equations in Eq.(\ref{lineareq}). Furthermore, the dilution technique~\cite{Foley:2005ac} can be applied to $\eta^{(t_{0})}_{[r]}$, which reduce the noise with small computational cost.

From the calculation for the $I=1$ $\pi\pi$ system~\cite{Akahoshi:ip}, we have found that the one-end trick technique is quite efficient for the analysis of hadron scatterings in the HAL QCD method with all-to-all propagators. We thus apply this as well to the kaon-nucleon system in this study.

 \nocite{*}
 
\bibliography{KNpreprint}


\end{document}